\documentclass[doublecol]{epl2} 

\begin{document}

\title{Parity-Time Symmetry meets Photonics: A New Twist in non-Hermitian Optics}
\shorttitle{Parity-Time Symmetry meets Photonics  ... } 

\author{Stefano Longhi \inst{1,2}}
\shortauthor{S. Longhi}

\institute{                    
  \inst{1}  Dipartimento di Fisica, Politecnico di Milano, Piazza L. da Vinci 32, I-20133 Milano, Italy\\
  \inst{2}  Istituto di Fotonica e Nanotecnlogie del Consiglio Nazionale delle Ricerche, sezione di Milano, Piazza L. da Vinci 32, I-20133 Milano, Italy}
 {Charge conjugation, parity, time reversal and other discrete symmetries}

\abstract{In the past decade, the concept of parity-time ($\mathcal{PT}$) symmetry, originally introduced in non-Hermitian extensions of quantum mechanical theories, has come into thinking of photonics, providing a fertile ground for studying, observing, and utilizing some of the peculiar aspects  of $\mathcal{PT}$ symmetry in optics. Together with related concepts of non-Hermitian physics of open quantum systems, such as non-Hermitian degeneracies (exceptional points) and spectral singularities, $\mathcal{PT}$ symmetry represents one among the most fruitful ideas introduced in optics in the past few years.
Judicious tailoring of optical gain and loss in integrated photonic structures has emerged as a new paradigm in shaping the flow of light in unprecedented ways, with major applications encompassing laser science and technology, optical sensing,  and optical material engineering.  In this perspective, I review some of the main achievements and emerging areas  of $\mathcal{PT}$-symmetric and non-Hermtian photonics, and provide an outline of challenges and directions for future research in one of the fastest growing research area of photonics.} 
\maketitle

\section{Introduction}
The search for new synthetic materials with desired optical properties and functionalities  has been one of the main driving forces of research in optics and photonics in the last few decades. Major examples of emerging synthetic optical media include photonic crystals and photonic crystal fibers, left-handed metamaterials, metal-dielectric (plasmonic) materials, epsilon-near-to-zero materials, metasurfaces, materials with topological and chiral properties, etc.  A recent  class of synthetic optical materials is that inspired by the concepts of parity ($\mathcal{P}$) and time reversal ($\mathcal{T}$) symmetries. In non-relativistic quantum mechanics, parity-time symmetry has attracted an increasing attention since  Carl Bender and collaborators suggested an hypothetical non-Hermitian extension of quantum mechanics \cite{r1,r2,r3,r4}, in which the postulate of Hermiticity of the underlying Hamiltonian is replaced by $\mathcal{PT}$ symmetric invariance. In the so-called unbroken $\mathcal{PT}$ phase, the Hamiltonian shows an entirely real energy spectrum in spite of being non-Hermitian. More general conditions for a non-Hermitian Hamiltonian to show an entirely real energy spectrum were also discussed later on \cite{r4bis}. While at a foundational level non-Hermitian extensions of quantum mechanics may pose some problems (for example superluminality \cite{r4tris}) and remain  controversial, effective non-Hermitian models are found in a wide variety of classical and quantum systems, for example in open quantum systems  \cite{r5,r6}, and the appearance of genuine non-Hermitian effects, such as those related to exceptional points (EPs) \cite{r7,r8}, are of broad physical relevance in quantum and classical physics. While in this work I will focus my attention to optics, one should mention that the concepts of $\mathcal{PT}$ symmetry and exceptional points have found a considerable interest in several other areas of physics, such as in electronic and microwave systems, mechanics, acoustics, atom optics and optomechanics. Earlier works on $\mathcal{PT}$ symmetry in optics were conceived to physical implement  scattering processes and phase transitions in $\mathcal{PT}$-symmetric systems  \cite{r9,r10,r11,r12,r13,r14,r14bis}. However, it was subsequently realized that $\mathcal{PT}$ symmetry can provide a fertile and technologically accessible tool to mold the flow of light in unprecedented ways with a wealth of promising applications. Earlier studies showed, for example, that asymmetric transport, anomalous diffraction and unidirectional invisibility can be realized in complex crystals at the $\mathcal{PT}$ symmetry breaking transition, where lattice bands start to merge \cite{r11,r15,r16,r16bis,r17,r18,r19,r20,r21,r21bis}. Invisibility, cloaking, non-Hermitian metamaterials and metasurfaces, based on $\mathcal{PT}$ symmetry concepts, were subsequently suggested \cite{r21bis,r21uff,r21tris,r21quatris,r21quintis}. $\mathcal{PT}$ symmetry has inspired the idea of a laser-absorber device \cite{r22,r23,r24}, i.e. the rather counterintuitive possibility to realize an optical device that simultaneously behaves like a laser (i.e. it is able to emit coherent light waves) and as a coherent absorber (i.e. it can fully annihilate coherent light waves incident onto it). Since then, applications of $\mathcal{PT}$ symmetry to laser science and technology, especially in integrated semiconductor laser devices, has seen a series or major achievements \cite{r25,r26,r27,r28,r29,r30,r31,r32,r33,r34,r35,r36}, such as single mode selection and laser stabilization \cite{r27,r28,r29,r31,r32,r33,r35}, polarization mode conversion \cite{r36}, and light structuring and transport \cite{r30,r34} to mention a few. Other emerging areas of research are those related to EPs, i.e. non-Hermitian degeneracies corresponding to the simultaneous coalescence of eigenenergies and corresponding eigenvectors. EPs show an intriguing chiral behavior arising from asymmetric breakdown of the adiabatic theorem \cite{r37,r38,r39} and find  application in optical sensing \cite{r40,r41,r42,r43}. Another important emerging area is the one on non-Hermitian topological lasers \cite{r45,r46,r47}. May be the great success of $\mathcal{PT}$ optics and non-Hermitian photonics  comes from the combined effect of a relatively simple theory with a rich physics behind, and the relative ease of fabrication complexity as compared to other synthetic materials (like photonic crystals and left-handed metamaterials). As a matter of fact, in few years $\mathcal{PT}$-optics  has turned from a simple theoretical curiosity into an important area of research in integrated photonics. Is everything in the field of $\mathcal{PT}$-optics and non-Hermitian photonics entirely new? One should acknowledge that some concepts of non-Hermitian physics and their implications in optics and laser science were partially known since many years. For example, the implications in optics of non-Hermitian degeneracies are know since quite a long time \cite{r8}, albeit they were not fully exploited in some interesting applications. Non-orthogonality of eigenmodes in laser theory, noticeably in unstable resonators  described by a highly non-Hermitian Hamiltonian,  is known to increase quantum noise by the so-called excess noise (or Petermann) factor \cite{r49,r50,r51} (the relation between $\mathcal{PT}$ symmetry and excess noise factor in a simple model is discussed in \cite{r52}). Non-normal dynamics in certain non-Hermitian laser models is known to give rise to transient growth, excitability and turbulence laser behavior \cite{r53,r54,r55}. Finally, loss engineering in semiconductor lasers was suggested and demonstrated in the early days of semiconductor lasers  to force single-mode operation (see the recent comment \cite{r56}). However, we can undoubtedly say that concepts like $\mathcal{PT}$ symmetry and non-Hermitian degeneracies are providing a major twist in the field of non-Hermitian photonics, with many of their potential applications yet to be explored.\par
In this perspective  I briefly review some of the simplest concepts underlying $\mathcal{PT}$-optics and non-Hermitian photonics, present a few emergent developments and applications in the field, and provide an outline of challenges and directions for future research. A huge number of papers have been published in this broad field in the past few years, and some reviews facing several aspects of $\mathcal{PT}$  symmetry in optics and beyond optics have been recently published \cite{r57,r58,r59,r60,r61,r62}, to which we refer the reader.

\section{Simple concepts in $\mathcal{PT}$-symmetric optics}
The simplest optical model that describes $\mathcal{PT}$ symmetry in optics, originally introduced in Ref.\cite{r11}, is provided by the optical Schr\"odinger equation \cite{r12}, which describes rather generally paraxial wave propagation of light waves  at wavelength $\lambda$ in an inhomogeneous dielectric medium with a $z$-independent transverse refractive index distribution $n(x)$, that slightly deviates from a reference (substrate) index $n_s$. The evolution of the electric field envelope $\psi(x,z)$ along the paraxial distance $z$ reads \cite{r11,r12}
\begin{equation}
i \hbar \partial_{z}\psi= -(\hbar^2)/(2n_s) \partial^2_x \psi +V(x) \psi \equiv \mathcal{H} \psi
\end{equation}
\begin{figure}
\onefigure[width=7.8cm]{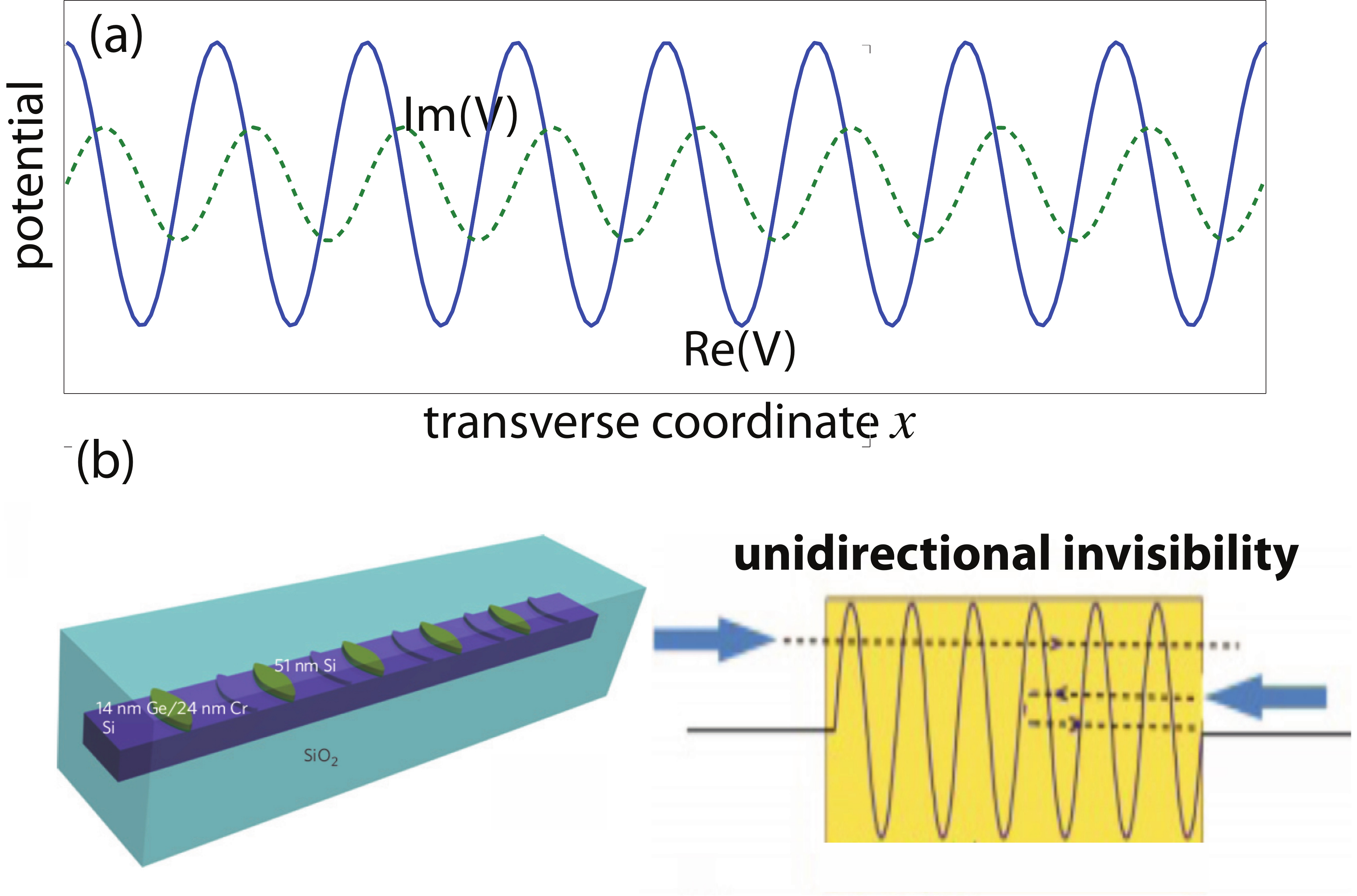}
\caption{ (a) Schematic of a $\mathcal{PT}$-symmetric sinusoidal potential. (b) Realization of the potential in a silicon waveguide based on combined index and loss gratings. Near the symmetry breaking point the crystal in unidirectionally invisible: waves propagating from left to right are not scattered off by the potential, while waves propagating from right to left are.}
\end{figure}
\begin{figure}
\onefigure[width=7.8cm]{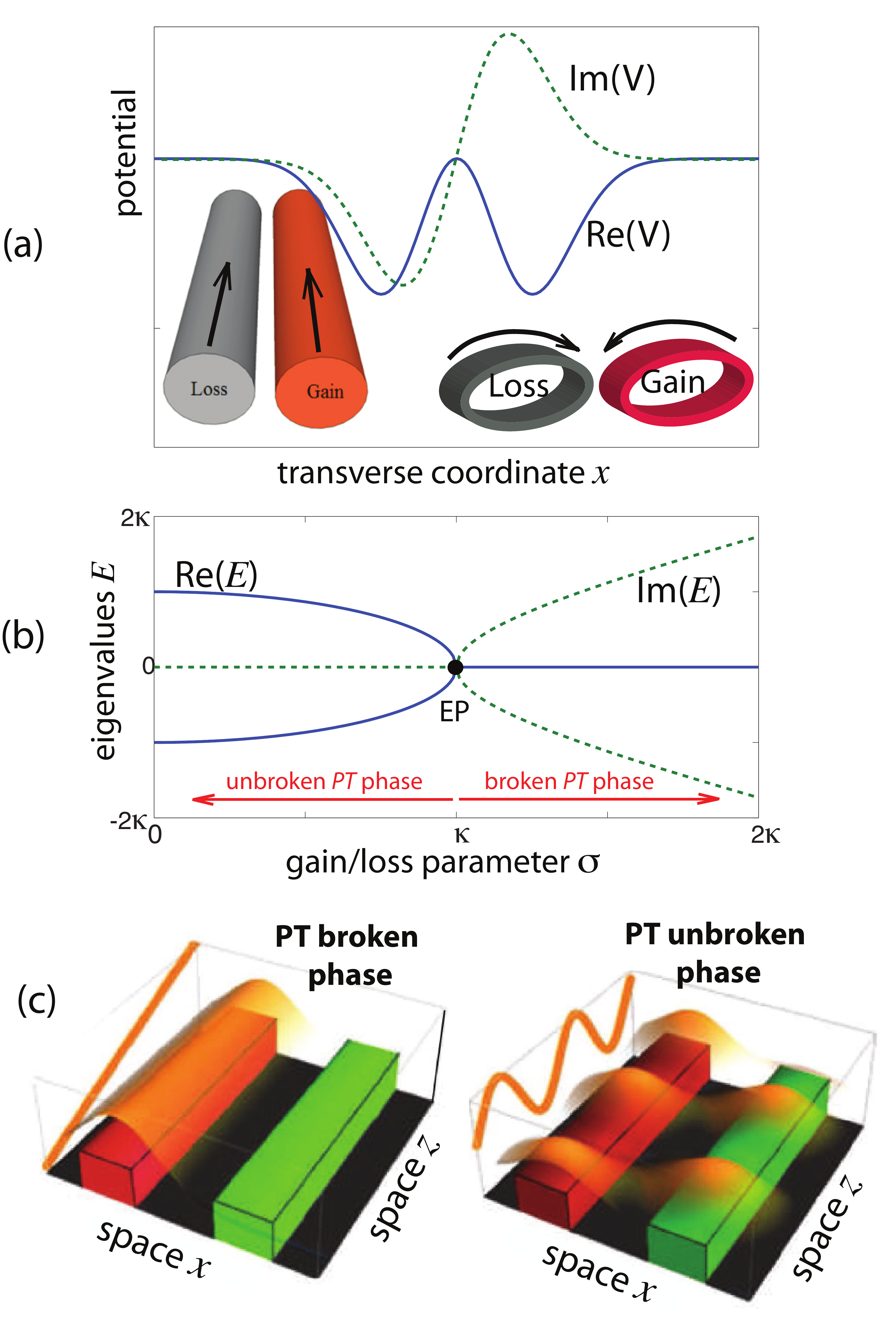}
\caption{ (a) Schematic of a $\mathcal{PT}$-symmetric double-well potential. Insets: optical realization based on  coupled optical waveguides (left) and coupled microring resonators (right) with balanced gain/loss. (b) Behavior of the energy spectrum of the Hamiltonian (2) versus the gain/loss parameter $\sigma$. The symmetry breaking point at $\sigma= \kappa$ corresponds to an exceptional point, with coalescence of both eigenvalues and eigenfunctions of the matrix $\mathcal{H}$. (c) Schematic of light propagation in a $\mathcal{PT}$-symmetric optical directional coupler above (left panel) and below (right panel) the symmetry breaking point.}
\end{figure}
where $\hbar \equiv \lambda / (2 \pi)$ and $V(x) \simeq n_s-n(x)$ as a definition of the optical potential.  Clearly, Eq.(1) is formally equivalent to the Schr\"odinger equation for a spineless quantum particle of mass $m=n_s$ in a potential $V(x)$, where the spatial evolution of the electric field along the paraxial distance $z$ emulates the temporal evolution of the quantum mechanical wave function. A non-Hermitian Hamiltonian $\mathcal{H}$ corresponds to a complex potential $V(x)$, i.e. to a complex refractive index distribution $n(x)$. As is well known, the imaginary part of the refractive index corresponds to optical gain or loss of the dielectric medium.  $\mathcal{H}$ is said to be $\mathcal{PT}$ symmetric if it commutes with the $\mathcal{PT}$ operator, i.e. $ [ \mathcal{H}, \mathcal{PT}]=0$ \cite{r1,r3}. For Eq.(1), $\mathcal{PT}$ symmetry implies $V(-x)=V^*(x)$, i.e. the real (imaginary) part of the potential should be an even (odd) function under space reflection $ x \rightarrow -x$. This means that a $\mathcal{PT}$ symmetric optical potential requires a balanced distribution of gain and loss in the medium. Since $\mathcal{PT}$ is not a linear operator, it is not ensured that the two operators $\mathcal{PT}$ and $\mathcal{H}$ share a common set of eigenfunctions and eigenvalues, even though they commute. If they share the same set of eigenfunctions, we say that the $\mathcal{PT}$ symmetry is unbroken and the energy spectrum of $\mathcal{H}$ is entirely real. Conversely, if the set of eigenfunctions of the two operators do not coincide, the $\mathcal{PT}$ symmetry is broken. In this case the energy spectrum of $\mathcal{H}$ is complex and energies are either real or appear in complex conjugate pairs. Generally, as the strength of the non-Hermitian part of the potential increases, a phase transition, from unbroken to broken $\mathcal{PT}$ phases, is observed. Let $V(x)=V_R(x)+i \sigma V_I(x)$, where $V_R$ and $\sigma V_I$ are the real (index) and imaginary (gain/loss) parts of the complex potential, and $\sigma$ a dimensionless parameter that measures the strength of the non-Hermitian part of the potential ($\sigma=0$ in the Hermitian limit). As a general rule of thumb, there is a critical value $\sigma=\sigma_c \geq 0$ such that the $\mathcal{PT}$ phase is unbroken for $\sigma \leq \sigma_c$ and broken for $\sigma> \sigma_c$. A special behavior is found at the phase transition point $\sigma=\sigma_c$, where the Hamiltonian $\mathcal{H}$ becomes defective owing to the appearance of EPs \cite{r6,r7} (for finite dimensional systems) or spectral singularities \cite{r14bis,r64,r65} (for infinite dimensional systems), i.e. the set of proper and improper eigenfunctions of $\mathcal{H}$ ceases to be a complete set.  Let us discuss two main examples that found interesting applications in optics. The first one is the complex sinusoidal (periodic) potential $V(x)=V_0[\cos(kx)+i \sigma \sin (kx)]$, which describes light scattering in a mixed index and gain/loss grating \cite{r11,r15,r16,r18,r66}; see Fig.1(a). In this case $\sigma_c=1$ and light scattering near the symmetry breaking point turns out to be unidirectional \cite{r15,r16}. This means that a wave propagating in one direction does not see any scattering potential, i.e. it freely propagates, while it is scattered off by the sinusoidal potential in the opposite direction [Fig.1(b)]. Probing the motion by an external force results, for example, in unidirectional Bloch oscillations \cite{r15,r19,r67,r68}. The second example is a double-well potential, with gain in one well and loss in the other one \cite{r69}, which can be realized by two coupled optical waveguides (a $\mathcal{PT}$-symmetric directional coupler) or two ring resonators, one active and the other lossy [Fig.2(a)]. Using coupled-mode theory,  the Hamiltonian of the $\mathcal{PT}$ coupler is simply described by the $2 \times 2$ Hamiltonian
\begin{equation}
\mathcal{H}= \left( 
\begin{array}{cc}
-i \sigma & \kappa \\
\kappa & i \sigma
\end{array}
\right)
\end{equation}
where $\kappa$ is the coupling constant and $\sigma$ the balanced gain/loss parameter in the two waveguides/rings. The eigenenergies of $\mathcal{H}$ are given by $E= \pm \sqrt{\kappa^2-\sigma^2}$, which are entirely real and distinct for $\sigma < \sigma_c$ with $\sigma_c \equiv \kappa$. At the symmetry breaking transition point $\sigma=\sigma_c=\kappa$, the two eigenvalues along with their corresponding eigenfunctions coalesce and an EP point arises. Finally, in the broken $\mathcal{PT}$ phase two pairs of complex conjugate eigenenergies are found [Fig.2(b)].  While for $\sigma < \sigma_c$ light oscillates between the two waveguides (Rabi-like oscillations), at and above the symmetry breaking transition a secular amplification of light in the waveguide with gain is observed [Fig.2(c)]. The $\mathcal{PT}$ coupler enabled the first observation in optics of $\mathcal{PT}$ symmetry breaking transition \cite{r13,r14}.\par
Beyond the paraxial and scalar theory, $\mathcal{PT}$ symmetry can be introduced for vectorial e.m. fields in general optical media within full Maxwell$^{\prime}$s equations \cite{r21uff}. In this framework, complex-coordinate transformation optics may be exploited for systematic generation, design, and modeling of a rather broad class of $\mathcal{PT}$-symmetric metamaterials for a variety of applications \cite{r21tris,r21quatris,r21quintis}.
\begin{figure}
\onefigure[width=8.4cm]{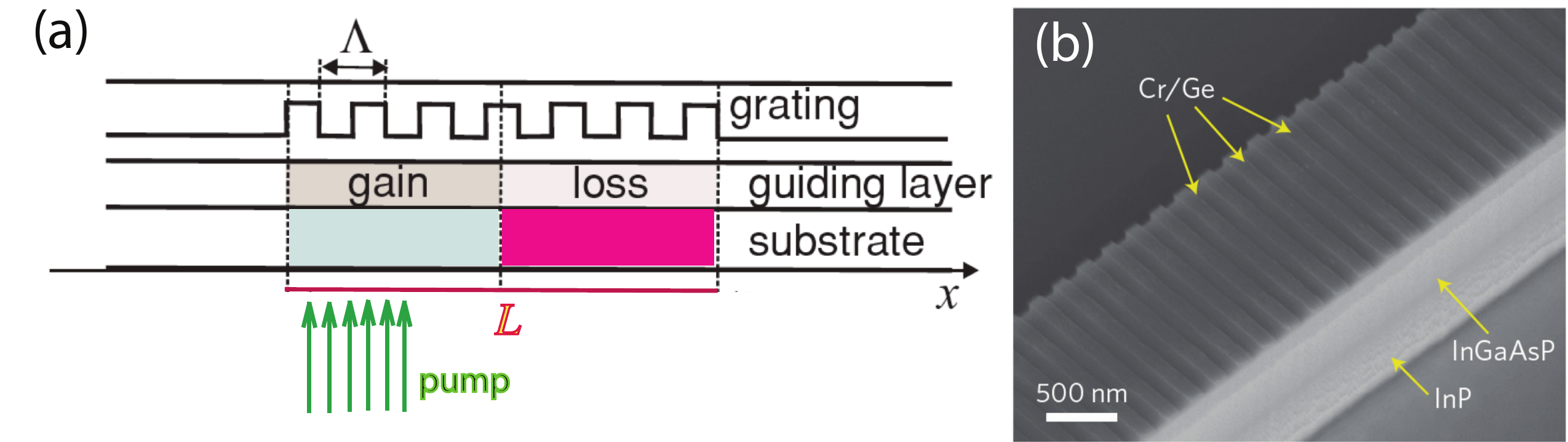}
\caption{ (a) Schematic of a $\mathcal{PT}$-symmetric laser-absorber in a distributed-feedback structure with balanced gain and loss regions \cite{r22}. (b) Realization of a laser-absorber device based on a a III-V semiconductor platform \cite{r24}.}
\end{figure}
\begin{figure}
\onefigure[width=8.2cm]{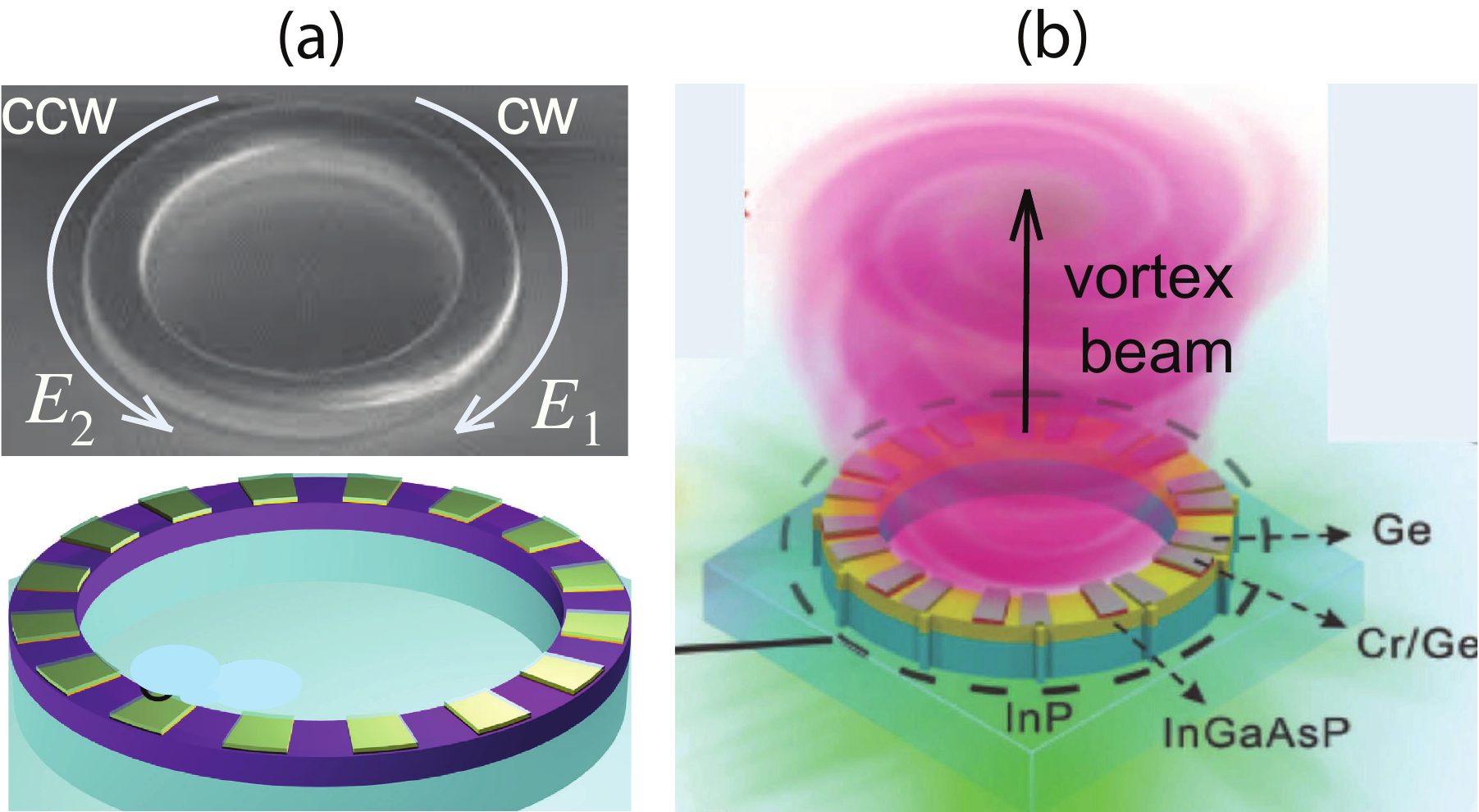}
\caption{ Unidirectional laser emission in a microring based on a $\mathcal{PT}$-symmetric grating. (a) In a microring two traveling wave modes, propagating clockwise (cw) or counterclockwise (ccw) with amplitudes $E_1$ and $E_2$, are degenerate and can compete via nonlinear laser dynamics. Depending on parameter conditions, unidirectionally bistable, bidirectional or oscillatory emission in the two modes can be observed. The use of a complex index/loss grating in a $\mathcal{PT}$ symmetric configuration (bottom panel) forces stable laser emission in one of the two traveling-wave modes. (b) The use of an additional grating can decouple light from the top of the microring in the form of a vortex laser beam carrying orbital angular momentum \cite{r34}.}
\end{figure}
\section{Some applications of $\mathcal{PT}$-symmetry in optics}
Here I provide a few examples, among many others, where $\mathcal{PT}$ symmetry finds applications in photonics.\par
{\it 1. Unidirectional invisibility in $\mathcal{PT}$ complex crystals.} As briefly mentioned above, light scattering in a complex periodic potential is asymmetric, and can completely vanish near the $\mathcal{PT}$ symmetry breaking transition for light propagation in one direction. Such an effect thus corresponds to unidirectional invisibility of the scattering medium \cite{r17,r18}. Within coupled-mode theory of shallow grating structures, such as effect was earlier predicted in Ref.\cite{r70}. From the experimental side, researchers have demonstrated unidirectional invisibility in complex crystals based on silicon waveguides \cite{r20} and in the temporal domain using a mesh of coupled fiber rings \cite{r19}. In the experimental setting of Ref.\cite{r20}, the real index and the imaginary loss modulations are spatially separated and can be controlled individually [Fig.1(b)]. The real index modulation is realized by either modulation of 
the waveguide width or by depositing additional dielectric materials (such as Ge or Si) on top of the waveguide. The phase-shifted loss grating is implemented by placing combined
dielectric-metal multilayered structures (such as a bilayer structure of Ge/Cr) on top of the waveguide.\par
{\it 2. Laser-absorber devices.} At the onset of lasing, in a simple one-dimensional semiclassical analysis laser oscillation, produced by an active medium inside an cavity of length $L$ occupying the spatial region $-L/2<x<L/2$, can be described by linear Mawxell$^{\prime}$s equations with a complex relative dielectric constant $\epsilon=\epsilon(x)$, where $\epsilon(x)=1$ outside the cavity, i.e. for $|x|>L/2$. The real and imaginary parts of the dielectric constants describe feedback provided by the mirrors and light amplification in the gain medium, respectively. If we write the electric field at frequency $\omega$ as $\mathcal{E}(x,t)=E(x) \exp(-i \omega t)+c.c.$, the spatial component $E(x)$ of the field satisfies the Helmholtz equation
\begin{equation}
\frac{d^2E}{dx^2}+\epsilon(x) \left( \frac{\omega}{c_0} \right)^2 E=0
\end{equation}
where $c_0$ is the speed of light in vacuum. For a laser with optical gain tuned at threshold, Eq.(3) admits of stationary outgoing wave solutions, i.e. solutions with the asymptotic form $E(x)=E_1 \exp(ikx)$ for $x>L/2$ and $E(x)=E_2 \exp(-ikx)$ for $x<-L/2$, where $k= \omega/c_0$ and $E_1$, $E_2$ are the amplitudes of the fields escaping from the cavity on the right and left sides, respectively. Clearly, $E^*(x)$ satisfies Eq.(3) provided that $\epsilon(x)$ is replaced by $\epsilon^*(x)$, i.e. optical gain is replaced by the same amount and spatial distribution of optical loss. Complex conjugation corresponds to time reversal $\mathcal{T}$, i.e. outgoing waves are transformed into incoming waves: this means that the time reversal of a laser, obtained by replacing the optical gain with the same amount of optical loss, realizes a perfect coherent absorber \cite{r71,r72,r73}. If we assume that $\epsilon^*(x)=\epsilon(-x)$,  i.e. if the laser system is $\mathcal{PT}$ symmetric, it readily follows that there should simultaneously exist, at the same real frequency $\omega$, a solution corresponding to outgoing waves and another one corresponding to incoming waves \cite{r22}. These are related one another by a space-time symmetry. In other words, a laser at threshold satisfying $\mathcal{PT}$ symmetry behaves as a coherent perfect absorber as well (frequency pulling effects might however induce a slight shift of frequency $\omega$ of lasing and absorbed waves). This is a rather counterintuitive effect, since it seems contradictory that an optical medium can amplify and fully absorb light simultaneously. As a matter of fact, time reversal symmetry ensures that an arbitrary lasing system, that could even emit irregular or chaotic light, has its own coherent perfect absorber \cite{r74}. An example of $\mathcal{PT}$ symmetric laser-absorber device, based on distributed optical feedback, was proposed in Ref.\cite{r22} [Fig.3(a)]. The experimental demonstration of a laser-absorber device has been recently reported in Ref.\cite{r24} using an active waveguide on a III-V semiconductor platform, with periodically placed bilayer Cr/Ge as the loss elements [Fig.3(b)]. The laser-absorber device provides a suitable platform for interferometric control of loss and amplification of light, with great potential to advance on-chip photonic modulation technologies.\par
{\it 3. Single-mode laser emission and structured light.} $\mathcal{PT}$ symmetry has inspired several methods of mode selection in integrated semiconductor lasers \cite{r26,r27,r28,r29,r30,r31,r32,r34,r35}. In particular, in a microring laser stable unidirectional laser emission can be forced by using a complex grating near the symmetry breaking point [Fig.4(a)]. This provides a rather unique means to obtain unidirectional oscillation without resorting to non-reciprocal elements, such as optical diodes, which is impossible in the micro/nano scale. The complex grating does not break time reversal symmetry, however the asymmetric feedback between clockwise and counter-clockwise modes induces laser dynamics to suppress one of the two circuiting modes. The method is robust against nonlinear instabilities that are known to set in in semiconductor lasers with slow carrier dynamics and a large linewidth enhancement factor \cite{r35}. An interesting application of unidirectional laser operation in a microring of few micrometer size is the ability to generate on an integrated optics platform structured light carrying orbital angular momentum (an optical vortex), which has been suggested and demonstrated in Ref.\cite{r34}; see Fig.4(b).

\par
{\it 4. Optical sensing and chirality with exceptional points.}
$\mathcal{PT}$ symmetry breaking in a finite dimensional system is usually related to the appearance of EPs \cite{r7,r8}, as illustrated in Fig.2(b). These are unique non-Hermitian degeneracies which show some intriguing properties that can be exploited in photonics. One key difference between exceptional points and Hermitian degeneracies is their sensitivity to perturbations. In a system operating around an Hermitian degeneracy, the resulting eigenvalue splitting is proportional to the perturbation strength $\epsilon$. Conversely, in a non-Hermitian system with a $N$-th order exceptional point, i.e. at which $N$ eigenenergies and corresponding eigenvectors coalesce, the splitting induced by the perturbation scales as $\sim \epsilon^{1/N}$ \cite{r40}. This results in an enhanced sensitivity of frequency splitting for a given strength of perturbations. Recent experiments have demonstrated high-sensitivity optical sensing using engineered high-quality optical microcavities that operate near an exceptional point \cite{r42,r43}. A target nanoscale object that enters the evanescent field of the cavity perturbs the system from its exceptional point, leading to frequency splitting with enhanced sensitivity. Another interesting property of exceptional point is chirality and asymmetric breakdown of the adiabatic theorem observed when an exceptional point is encircled. Let us consider a non-Hermitian Hamiltonian, like the one given by Eq.(2), dependent on two real parameters. The parameters are slowly varied in time, describing a closed path in phase space. An interesting case is the one where the path encircles an exceptional point. In this case, for extremely slow motion,  the system, initially prepared in an instantaneous eigenstate, evolves adiabatically remaining in the instantaneous eigenstate of the Hamiltonian. Owing to the topological nature of the branch point, after one cycle a state-flip is obtained \cite{r75}. If the motion is reversed, the system returns to its initial state. However, breakdown of adiabaticity can be observed in one circulation direction, but not in the opposite one, whenever the exceptional point is dynamically encircled not quasi statically \cite{r76,r77}. This introduces a chirality in the system, which has been observed in recent experiments \cite{r37,r38}. Chirality and asymmetric breakdown of the adiabatic theorem, observed when dynamically encircling an exceptional point, are rather general effects of non-Hermitian dynamics that do not necessarily require to encircle an EP. Chirality can be explained in terms of asymmetric transition rates for positive- or negative-frequency components of the time-varying part of the Hamiltonian \cite{r39}. Interestingly, the concept of Floquet EPs and dynamically-induced chirality, i.e. the creation of EPs by time-periodic modulation of a non-Hermitian Hamiltonian, has been  recently introduced in Ref.\cite{r78}, opening up possibilities to engineer EPs by periodic drivings. 

\section{Non-Hermitian photonics}
$\mathcal{PT}$ symmetry represents an important concept in recent developments of non-Hermitian photonics. However, a broader class of effects arising from loss and/or gain engineering of optical media can be observed without resorting to $\mathcal{PT}$ symmetry. Here I provide for the sake of illustration a few examples of recent advances in non-Hermitian photonics based on non-$\mathcal{PT}$-symmetric media.\par
{\it 1. Kramers-Kronig optical media.} Kramers-Kronig optical media represent a recent class of synthetic optical media with tailored index and loss/gain regions, which have been introduced by S.A.R. Horsely and coworkers in Ref.\cite{r79}. In such media, $\mathcal{PT}$ symmetry is replaced by the property that the real and imaginary parts of the dielectric permittivity are related one another by spatial Kramers-Kronig relations (Hilbert transform). A fascinating property of such media is that they are one-way or bidirectionally reflectionless, whatever the angle of incidence. Such a property, besides extending our comprehension of the fundamental phenomenon of reflection, may offer new ways for the design of antireflection surfaces and thin materials with efficient light absorption \cite{r80,r81}. Ongoing research in this area can be found in Refs.\cite{r82,r83,r84,r85,r86,r87,r87bis}.  \par
\begin{figure}
\onefigure[width=8.4cm]{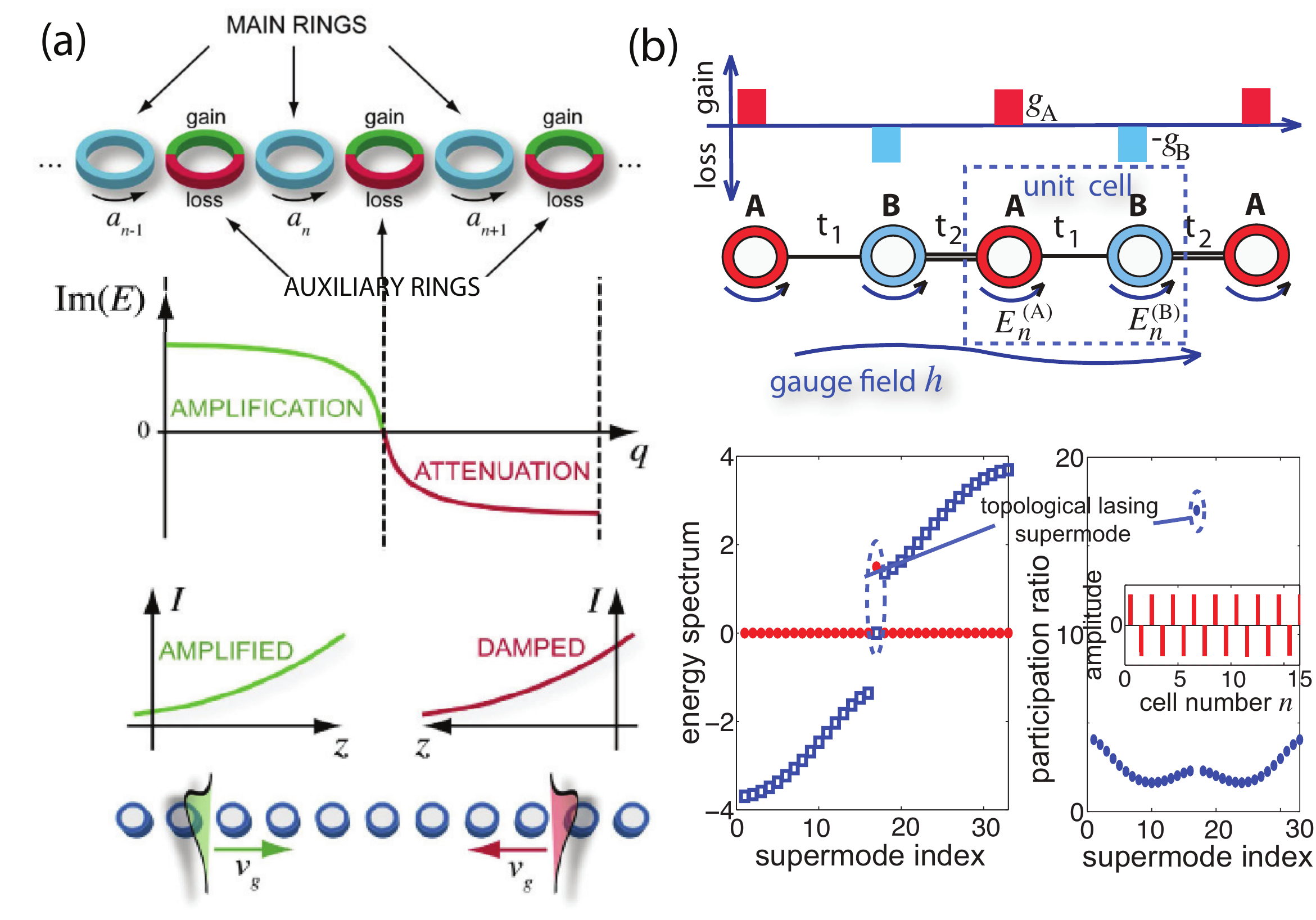}
\caption{ (a) Chain of coupled microring resonators with a synthetic imaginary gauge field realized using auxiliary rings in antiresonance with gain and loss \cite{r30}. The energy dispersion relation of the non-Hermitian lattice ensures unidirectional robust light transport, insensitive to disorder. (b) Topological laser array based on a Su-Schrieffer-Heeger microring chain with an imaginary gauge field \cite{r47}.}
\end{figure}
{\it 2. Imaginary gauge fields.} Gauge fields play an important role in topological photonics \cite{r88,r89,r90}. In tight-binding lattice models, i.e. within coupled-mode theory formalism, synthetic gauge fields are introduced by proper control of the phase $\phi$ of the coupling constants (Peierls$^{\prime}$ phase) between micro resonators or waveguides \cite{r89,r90,r91}. In Hermitian models, the Peierls$^{\prime}$ phase is real. However, in 1996 Hatano and Nelson introduced the idea of an {\it imaginary} gauge field to study non-Hermitian delocalization transition in the Anderson model \cite{r92}. They considered the one-dimensional tight-binding lattice Hamiltonian
\begin{equation}
\mathcal{H}  =   \sum_n \left[  \kappa \exp(i \phi )\hat {c}^{\dag}_nc_{n+1}+\kappa \exp(-i \phi) \hat{c}^{\dag}_{n+1} \hat{c}_n + V_n \hat{c}^{\dag}_n \hat{c}_n  \right]
\end{equation}
 where $\kappa$ is the hopping rate between adjacent sites, $V_n$ is the site energy that accounts for possible disorder in the lattice, and $\phi=i h$ is the imaginary gauge phase. The Hamiltonian $\mathcal{H}$ is clearly non-Hermitian owing to the imaginary gauge field. An important property of the  Hatano-Nelson Hamiltonian is to enable a directional transport along the lattice, which is robust against disorder in the chain \cite{r30,r93}; see Fig.5(a). This is related to the delocalization transition of eigenvectors induced by the imaginary gauge field \cite{r92}. A synthetic imaginary gauge field for photons can be realized using coupled microring resonators, with auxiliary anti-resonant rings with optical loss and gain, as shown in Fig.5(a) \cite{r30}. Besides of realizing robust photonic transport \cite{r30,r93}, in optics imaginary gauge fields can be useful for the design of topological lasers, as suggested in\cite{r47}. In \cite{r47}, a Su-Schrieffer-Heeger chain with imaginary gauge field is realized by a chain of coupled microring resonators [Fig.5(b)]. The imaginary  gauge field stretches all supermodes of the laser array at one edge of the chain, whereas the topologically-protected edge state of the Su-Schrieffer-Heeger lattice becomes delocalized in the chain. In this way, stable laser emission can be realized in a single supermode of the lattice [Fig.5(b)]. \par
 {\it 3. Supersymmetric non-Hermitian photonics.} Other kind of symmetries, known in quantum physics, can find applications in photonics. A noticeable example is provided by supersymmetric quantum mechanics, which can be extended to non-Hermitian Hamiltonians. For example, invisibility of discretized light propagation in waveguide lattices with gain and loss regions in a non-$\mathcal{PT}$-symmetric configuration, synthesized by supersymmetric quantum mechanics, has been suggested in Ref.\cite{r94}, whereas supermode selection in laser arrays based on supersymmetry has been proposed in \cite{r95}.  
 \par
\section{Emerging areas} $\mathcal{PT}$ optics is a rapidly growing field with several ramifications and developments. I would like just to mention a few emerging areas of research.\\
{\it $\mathcal{PT}$-symmetry in nonlinear optical systems.} Nonlinear systems in $\mathcal{PT}$ symmetric configurations are attracting a considerable attention. A wide variety of novel effects, which have no counterparts in traditional dissipative systems, are found in nonlinear $\mathcal{PT}$ symmetric models, such as stabilization of nonlinear states above the symmetry breaking point, symmetry breaking of nonlinear modes, and peculiar soliton dynamics. Recent reviews on nonlinear $\mathcal{PT}$-symmetric systems, also beyond optics, can be found in \cite{r59,r60}.\par
{\it $\mathcal{PT}$-symmetry in plasmonic and metamaterial structures.} The concept of $\mathcal{PT}$ symmetry in dielectric optical media has inspired new design criteria in other optical media such as in plasmonic, metamaterial and epsilon-near-to-zero structures. Ongoing research in this area can be found in \cite{r21bis,r21uff,r21tris,r21quatris,r21quintis,palle1,palle2,palle3}. Since the technological realization of such structures is more challenging than $\mathcal{PT}$-symmetric dielectric media, experiments and device applications in this area are still to come. \par
{\it Topological non-Hermitian photonics.}  The meeting between topological photonics and non-Hermitian photonics is a promising research area which is expected to provide major advances in both theoretical and applied aspects of non-Hermitian photonics \cite{r96,r97,r98,r99,r100,r101,r102,r103}. The ideas of topology have found tremendous success in Hermitian photonic systems. However, even richer properties are expected to arise when dealing with non-Hermitian systems. To what extent concepts of topological matter, elaborated within Hermitian models, can be extended and find application in non-Hermitian systems remains a rather unexplored and fascinating area of research.  Recent works have shown, for example, formation of non-Hermitian topological edge states and interface states via quantum phase transitions \cite{r97,r98,r99,r101}, topological lasing in non-Hermitian photonic structures \cite{r45,r46,r47}, and formation of Fermi arc and polarization half charge via EPs in photonic crystal structures \cite{r103}.

\section{ Conclusions}
The concept of $\mathcal{PT}$ symmetry entered into optics about one decade ago \cite{r9,r10,r11}, when some authors realized that light scattering in certain optical structures with balanced optical gain and loss can provide a physical realization of $\mathcal{PT}$-symmetric Hamiltonians introduced by Carl Bender in the framework of a non-Hermitian extension of quantum mechanics \cite{r1}.  Rather surprising, such a concept has proven to be very fruitful in several area of photonics, with a wealth of applications ranging from laser technology to optical sensing and material engineering. Together with other concepts borrowed from the physics of non-Hermitian systems, such as EPs, $\mathcal{PT}$ symmetry has provided great inspiration in the design of new synthetic optical media with unprecedented functionalities compatible with current semiconductor technologies. As a matter of fact, non-Hermitian photonics is a fastest growing field of research with several ramifications and applications yet to come. Non-Hermitian topological photonics, $\mathcal{PT}$ symmetry in metamaterials, metasurfaces and plasmonic systems, and $\mathcal{PT}$-symmetric optomechanics are a few emerging research areas which promise major advances in the science and technology of light.

\end{document}